\documentstyle[preprint,aps,epsf]{revtex} 
\newcommand{\be}[1]{\begin{equation}\label{#1}}
\newcommand{\ee}{\end{equation}}     
\newcommand{\bea}{\begin{eqnarray}}
\newcommand{\eea}{\end{eqnarray}} 
\newcommand{\eq}[1]{Eq.\ (\ref{#1})}
\newcommand{\fig}[1]{Fig.\ (\ref{#1})}
\newcommand{\pic}[2]{\epsfxsize #1 cm
\epsffile{sand-hyd#2.eps}}    
\begin{document}  
\tighten   
\draft 
\title{\bf  Semiclassical time--dependent propagation in three 
dimensions: How accurate is it for a Coulomb potential?}
\author{Gerd van de Sand and  Jan M. Rost}
\address{-- Theoretical Quantum Dynamics --\\
Fakult\"at f\"ur Physik, Universit\"at Freiburg,
Hermann--Herder--Str.  3, D--79104 Freiburg,
Germany}
\date{\today}
\maketitle
\begin{abstract} 
A unified semiclassical time propagator is used to calculate the 
semiclassical time-correlation function in three cartesian dimensions
for  a particle moving in an attractive Coulomb potential. It is 
demonstrated that under these conditions the singularity of the 
potential does not cause any difficulties and the Coulomb interaction 
can be treated as any other non-singular potential. Moreover, by 
virtue of our three-dimensional calculation, we can explain the 
discrepancies between  previous semiclassical and quantum results 
obtained for the one-dimensional radial Coulomb problem.
\end{abstract}
\draft

Semiclassical propagation in time has been studied intensively 
in two dimensions \cite{Kay94b,Gro96,Mil96,Tan96}. There are by far not as many 
applications to higher dimensional problems, in particular not 
in connection with the singular Coulomb potential. Our motivation for this 
study is three-fold: Firstly to see, if the advanced semiclassical 
propagation  techniques in time, namely the Herman-Kluk propagator \cite{HK84,Kay94a,Gro98},
can be implemented for realistic problems of scattering theory 
involving long range forces. Secondly to see, if we can 
avoid to regularize the Coulomb singularity 
in the classical equations of motion if we work in three (cartesian) 
dimensions, and thirdly, to clarify the reason for the small, but pertinent 
discrepancies with the quantum result in two previous, one--dimensional
semiclassical calculations of the hydrogen spectrum from the time domain \cite{Tom93,Ezr94}. 
As it will turn out, the Coulomb problem with the Hamiltonian 
(we work in atomic units unless stated otherwise)
\be{hamil}
 H = \frac{{\bf p}^{2}}{2} + \frac{Z}{\left| {\bf r} \right| }
\ee
can be propagated in time 
semiclassically without taking any special care of the singularity in 
the  potential which  poses a lot of difficulties for the 
one-dimensional radial problem if $Z<0$, i.e., if the potential is 
attractive as in the case of hydrogen $(Z=-1)$ which we take as an 
example in the following.  
The relevant information in the time domain is the autocorrelation function
\be{corr}
c(t) =  \langle\psi|K|\psi\rangle
\ee
where 
\be{prop}
K({\bf r}, {\bf r}',t) = \langle {\bf r} |e^{-iHt/\hbar}| { \bf r}' \rangle
\ee
is the propagator in the coordinate representation.
By diagonalizing  $K$ in \eq{corr}  one can express the 
autocorrelation function  with the time evolution 
operator $U(t)$,
\be{corr2}
c(t) = \langle \psi(0)|U(t)|\psi(0)\rangle \equiv \langle \psi(0)|\psi(t)\rangle.
\ee
This form has the obvious interpretation of correlating the time 
evolving wavefunction $\psi(t)$ at each time with its value at $t=0$. 
The extraction of the energy spectrum  from \eq{corr2} is  routinely 
performed by Fourier transform,
\be{power}
\sigma(\omega) = \int c(t) \, e^{i\omega t} \, dt.
\ee
Expanding formally  $\psi(t)$ in terms of eigenfunctions
\be{wavet}
\psi(t) = \sum_{nlm} a_{nlm} \, \Phi_{nlm} \, e^{iE_{n}t/\hbar}
\ee
and inserting \eq{wavet} into \eq{power}
one sees that 
\be{powspec}
\sigma(\omega) = \sum_{n}\delta(\omega - E_{n}/\hbar) \, b_{n}.
\ee
Hence, the power spectrum $\sigma(\omega)$ exhibits peaks at the 
eigenenergies  of the system with weights given by 
\be{weight}
b_{n}=\sum_{lm} | a_{nlm} |^2 \equiv \sum_{lm} \left| \langle\psi(0)|\Phi_{nlm} \rangle \right| ^2
\ee
which are determined by the overlap with the initial wavepacket $\psi(0)$. 
For a finite  propagation time $t$ the peaks will have a 
finite width $\Gamma$.  Based on an idea by Neuhauser and Wall \cite{WN95} 
Mandelshtam and Taylor \cite{MT97} have devised 
the so called filter--diagonalization method as an alternative to the 
Fourier transform for extracting the energy spectrum from a 
finite time signal $c(t)$. 
Assuming a form 
\be{corr3}
c(t) = \sum_{j}a_{j} \, e^{iE_{j}t/\hbar}
\ee
with $a_{j}$ and $E_{j}$ being complex the filter--diagonalization 
allows one to extract $E_{n}$ and $b_{n}$ directly from 
\eq{powspec}. We will use this stable and accurate method to obtain
the spectral information from the time signal  $c(t)$ which has been 
calculated semiclassically as follows.
For the initial wavefunction we have taken a normalized Gaussian wavepacket,
$\psi(0)=(\gamma^{2}/\pi)^{3/4}f_{\gamma}({\bf r}, {\bf r}_0, {\bf p}_0
) $ with 
\be{psii}
f_{\gamma}({\bf r},{\bf r}_0, {\bf p}_0) = 
{\rm exp} \left( -\frac{\gamma^{2}}{2}({\bf r}- {\bf r}_0)^2 + \frac{i}{\hbar} {\bf p}_0 ({\bf r} - {\bf r}_0) \right)
\ee
where the inverse width $\gamma$ of the wavepacket and its center 
$({\bf p}_0,{\bf r}_0)$ in phase space determine with which weight the hydrogenic eigenfunctions 
are covered  by $\psi(0)$, see \eq{weight}.
 
The semiclassical propagator according to   Herman and Kluk \cite{HK84} 
is formulated as an integral over phase space, 
\bea \label{HK}
K_{\gamma}({\bf r},{\bf r}',t) & = & \frac{1}{(2 \pi \hbar)^3} \int \! \! \! 
\int \! d^3{\bf q} \, d^3 {\bf p} \,\,  
R_{\gamma}({\bf p}_t, {\bf q}_t) \,\, {\rm exp} \left( \frac{i}{\hbar} S \left( {\bf p}_t, {\bf q}_t \right) \right)  \nonumber \\
& & f_{\gamma} \left( {\bf r},{\bf q}_t,{\bf p}_{t} \right)
f^{*}_{\gamma} \left( {\bf r}',{\bf q},{\bf p} \right),
\eea
where ${\bf q}_t = {\bf q}(t)$ and ${\bf p}_t = {\bf p}(t)$ are the  phase 
space values at time $t$ of the trajectory started at time $t=0$ with $({\bf q},{\bf p})$
and propagated under the classical Hamiltonian \eq{hamil}.
The action $S = \int (p\dot q-H) \, dt$ accumulated along the trajectory 
enters \eq{HK}  as well as  the probability density of each trajectory 
$R_{\gamma}({\bf p}_t, {\bf q}_t)$  which 
contains   all four blocks  $M_{ij}$ of the monodromy matrix,
\be{mono}
\left( \begin{array}{c}
\delta {\bf q}_t \\ \delta {\bf p}_t \end{array} \right) =
\left( \begin{array}{cc}
M_{qq}(t) & M_{qp}(t) \\ M_{pq}(t) & M_{pp}(t) \end{array} \right) \,   
\left( \begin{array}{c}
\delta {\bf q} \\ \delta {\bf p} \end{array} \right) \, .
\ee
The actual form of  the probability density depends on a width 
parameter $\gamma$ which determines the admixture of the different 
blocks $M_{ij}$, 
\be{monofak}
R_{\gamma}({\bf p}_t, {\bf q}_t) = \left| \frac{1}{2} \left( M_{qq} + 
M_{pp} - i \gamma^2 \, \hbar M_{qp} - \frac{1}{i \gamma^2 \hbar} 
 \, M_{pq} \right) \right| ^{1/2}.
\ee
Although this semiclassical propagator is not uniquely defined through 
its dependence from a suitable chosen parameter $\gamma$ it has several 
important advantages over other forms. Firstly, it is 
globally uniformized  since at a caustic 
$R_{\gamma}$ remains always finite. Secondly, and this is of 
considerable  relevance for practical applications, one does not have 
to keep track of Maslov indices.
Instead one has to make $R_{\gamma}$ continuous as the radicant crosses the branch cut. 
Inserting \eq{HK} and \eq{psii} into \eq{corr} we obtain a 
particularly simple form for the 
semiclassical correlation function if the width of the initial 
Gaussian in $\psi(0)$ and of the propagator itself in $R_{\gamma}$ 
are chosen to be the same, 
\be{semcorr}
c_{sc}(t)  =  \frac{1}{(2 \pi \hbar)^3} \int \! \! \! \int \! d^3{\bf q} \, d^3 {\bf p} \, \, 
R({\bf p}_t, {\bf q}_t) \nonumber \, \, {\rm exp} \left(\frac{i}{\hbar} S({\bf p}_t, {\bf q}_t) \right) \,
g_\gamma \left( {\bf q},{\bf p}, {\bf r}_0, {\bf p}_0 \right) \, 
g_{\gamma}^{*} \left( {\bf q}_t,{\bf p}_t, {\bf r}_0, {\bf p}_0 \right) \, ,
\ee
where
\be{gauss}
g_\gamma \left( {\bf q},{\bf p}, {\bf q}', {\bf p}' \right) = 
{\rm exp} \left( -\frac{\gamma^2}{4}({\bf q} - {\bf q}')^2 - \frac{1}{4 \gamma^2} 
({\bf p} - {\bf p}')^2 + \frac{i}{2 \hbar} 
({\bf p} + {\bf p}')({\bf q} - {\bf q}') \right) 
\ee

The integrations over ${\bf r}$ and ${\bf r}'$  have  been 
carried out  analytically  which is possible due to 
our  choice of the initial wavepacket as a Gaussian.
The remaining integral in \eq{semcorr} is over the  entire phase space and  
in practice $c_{sc}(t)$ is calculated  by 
Monte Carlo integration where each randomly chosen phase space point
$({\bf q},{\bf p})$ represents the initial conditions for a classical 
trajectory.  It evolves in time under Hamilton's equations generated 
by the  
Hamiltonian of \eq{hamil}  and with the values $({\bf q}_t, {\bf p}_t)$ 
entering \eq{semcorr}. The number of sampling points (trajectories to 
be run)  to achieve convergence depends  very much on the initial 
wavepacket $\psi(0)$. In general it varies between a couple of thousand 
and a couple of million trajectories. 

Our first objective is to compare our results with earlier one 
dimensional calculations \cite{Tom93,Ezr94}. Although we work in three 
dimensions we can mimic the one dimensional results to some extent by choosing a 
similar initial wavepacket.

\begin{figure}
\pic{9}{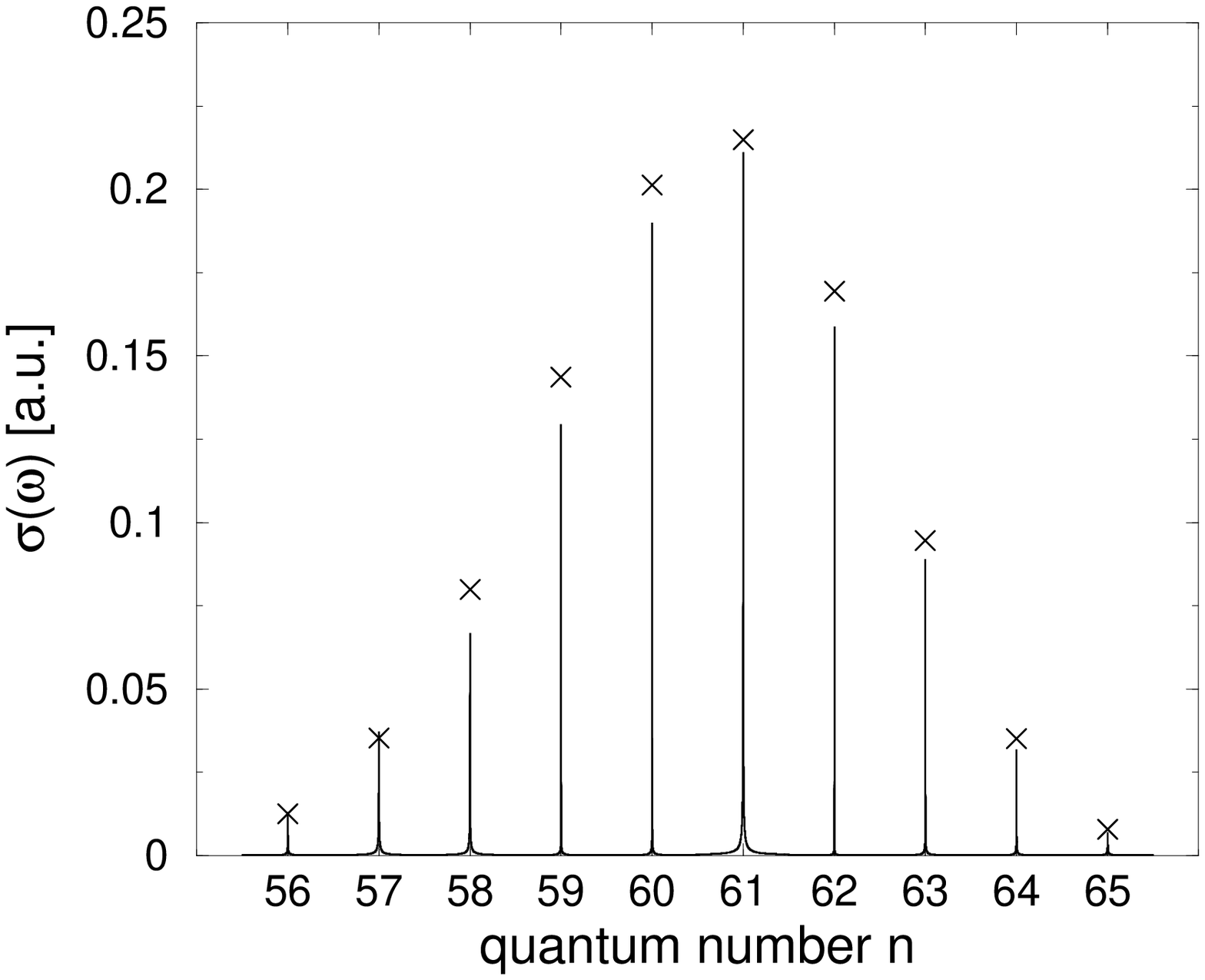}
\caption[]{Semiclassical spectrum (full curve) compared to quantum spectrum
(crosses) with parameters 
${\bf r}_0 = (0, 0, 6000 \, a.u.)$, ${\bf p}_0 = (0, 0, 0)$ and $\gamma^2 = 2/600^2$.} \label{swave}
\end{figure}

The result with parameters similar to those 
from \cite{Tom93} is shown in \fig{swave}.
One sees excellent agreement concerning the positions of the
peaks with quantum mechanics (crosses) and small but noticeable
deviations of the weights of the states. This observation, as the 
entire figure, is very similar to the  findings of \cite{Tom93} and 
\cite{Ezr94}.  However, we would like to emphasize that our result has been 
obtained from a `routinely' applied semiclassical propagator without 
explicit regularization or  Langer corrections or any other means 
implemented to deal with the Coulomb singularity. 
These complication, dealt with in \cite{Ezr94}, occur only if one uses 
explicitly curved linear coordinates where the problem of  the order
of operators renders the classical--quantum correspondence difficult.
This becomes obvious if the semiclassical propagator is derived from  
Feynman's path integral, see e.g., Kleinert's book on path integrals 
\cite{Klei95}. 
Of course,  the prize one has to pay in order to avoid these 
complications in, e.g.,  a radial coordinate, is to work in a higher
dimensional (cartesian) space as it has been done here.  

However, even
in our approach we should regularize trajectories which hit the 
Coulomb
singularity directly (impact parameter zero). Fortunately, these 'head
on' trajectories are of measure zero among all trajectories contained 
in the initial conditions and with a Monte Carlo method they are
hardly ever encountered.  Even if  such a trajectory is selected by 
chance, one can safely discard its contribution to the propagator.

\begin{figure}
\pic{9}{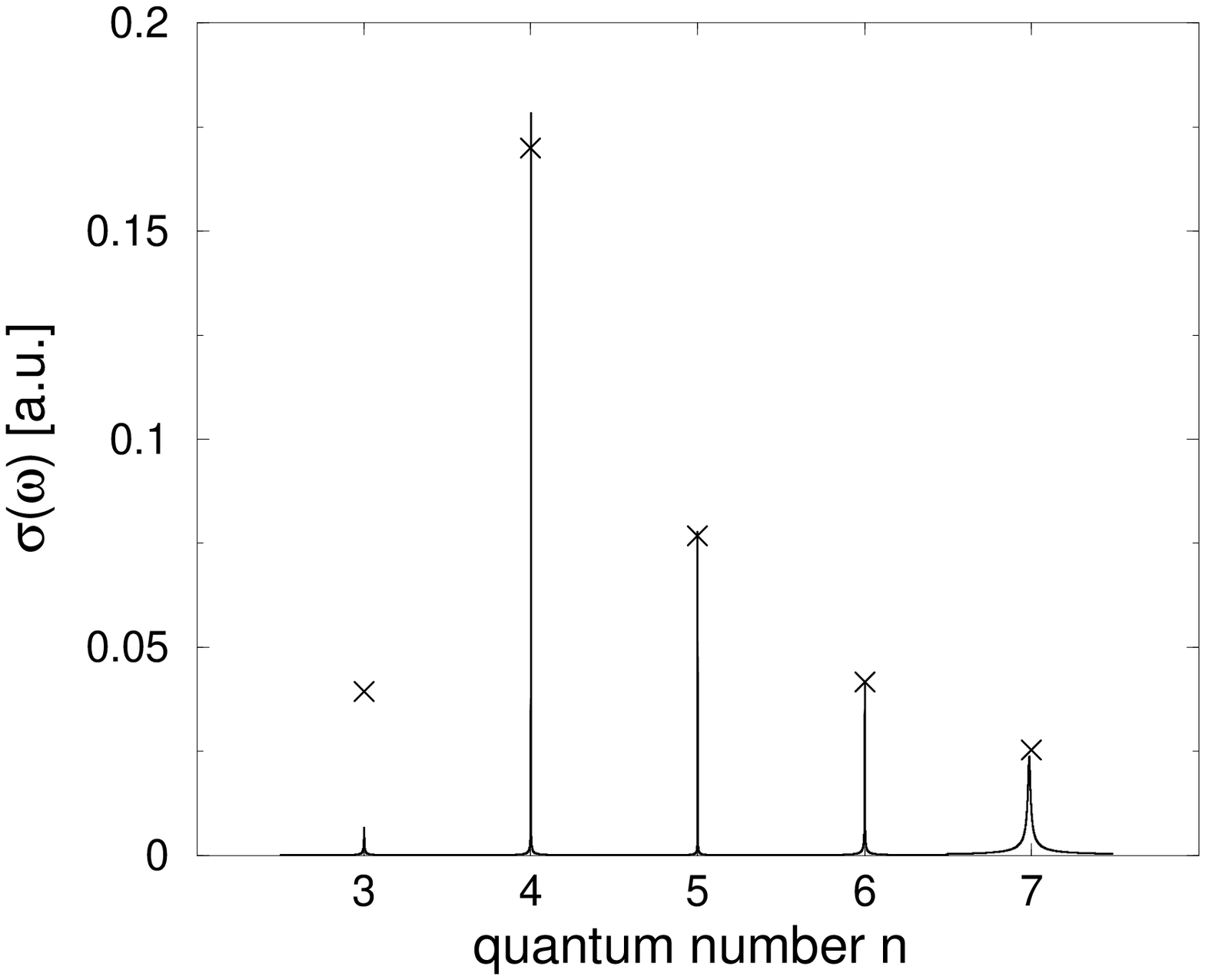}
\caption[]{Semiclassical spectrum (full curve) compared to quantum spectrum
(crosses) with parameters 
${\bf r}_0 = (0, 0, 20 \, a.u.)$, ${\bf p}_0 = (0, 0, 0)$ and $\gamma^2 = 0.1$.} 
\label{low}
\end{figure}

\begin{figure}
\pic{9}{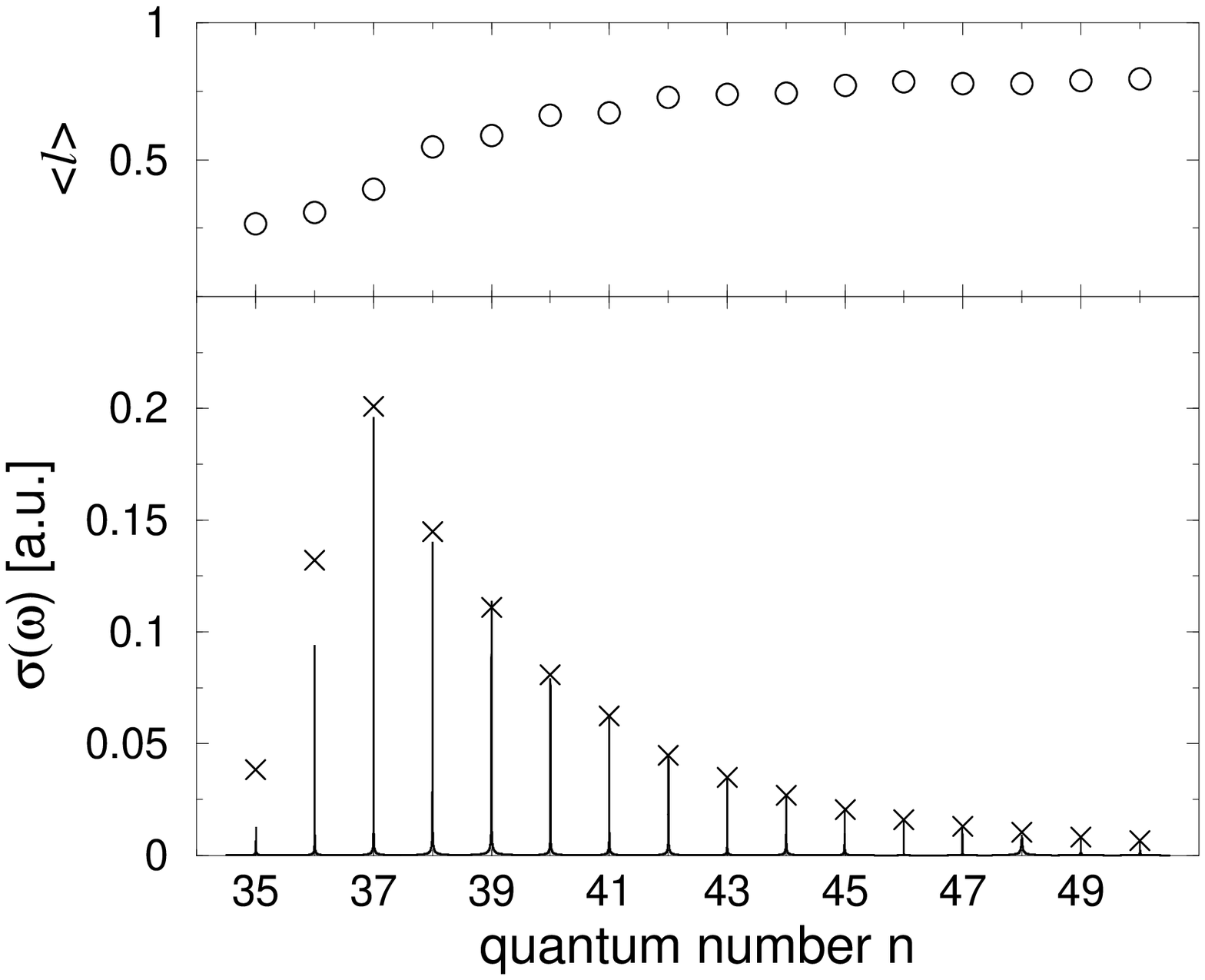}
\caption[]{Lower panel: Semiclassical spectrum (full curve) compared to
quantum spectrum (crosses) with parameters 
${\bf r}_0 = (0, 0, 2500 \, a.u.)$, ${\bf p}_0 = (0, 0, 0)$ and $\gamma^2 = 0.0001$; upper panel: average
normalized angular momentum according to \eq{lnorm} (circles).} \label{medium}
\end{figure}

The direct semiclassical integration is in principle able to 
reproduce the spectrum even for low excitation as can be seen in 
 \fig{low}, and, less surprisingly, for medium excitation 
(\fig{medium}). However, a systematic trend is apparent from these two 
spectra: The agreement of the weights is much better to the left of 
the largest peak than to the right. To understand this effect we have
plotted in \fig{medium} the average  angular momentum fraction
\be{lnorm}
<l> \, = \, \frac{1}{n-1}
\frac{\sum_{lm} \, l \, \left| a_{nlm} \right|^2}{\sum_{lm} \left| a_{nlm} \right|^2} \, ,
\ee
contained in the weights $b_{n}$ in addition.
One sees that good agreement goes 
along with a large fraction of high angular momentum states in the 
initial wavepacket and vice versa. 

\begin{figure}
\pic{9}{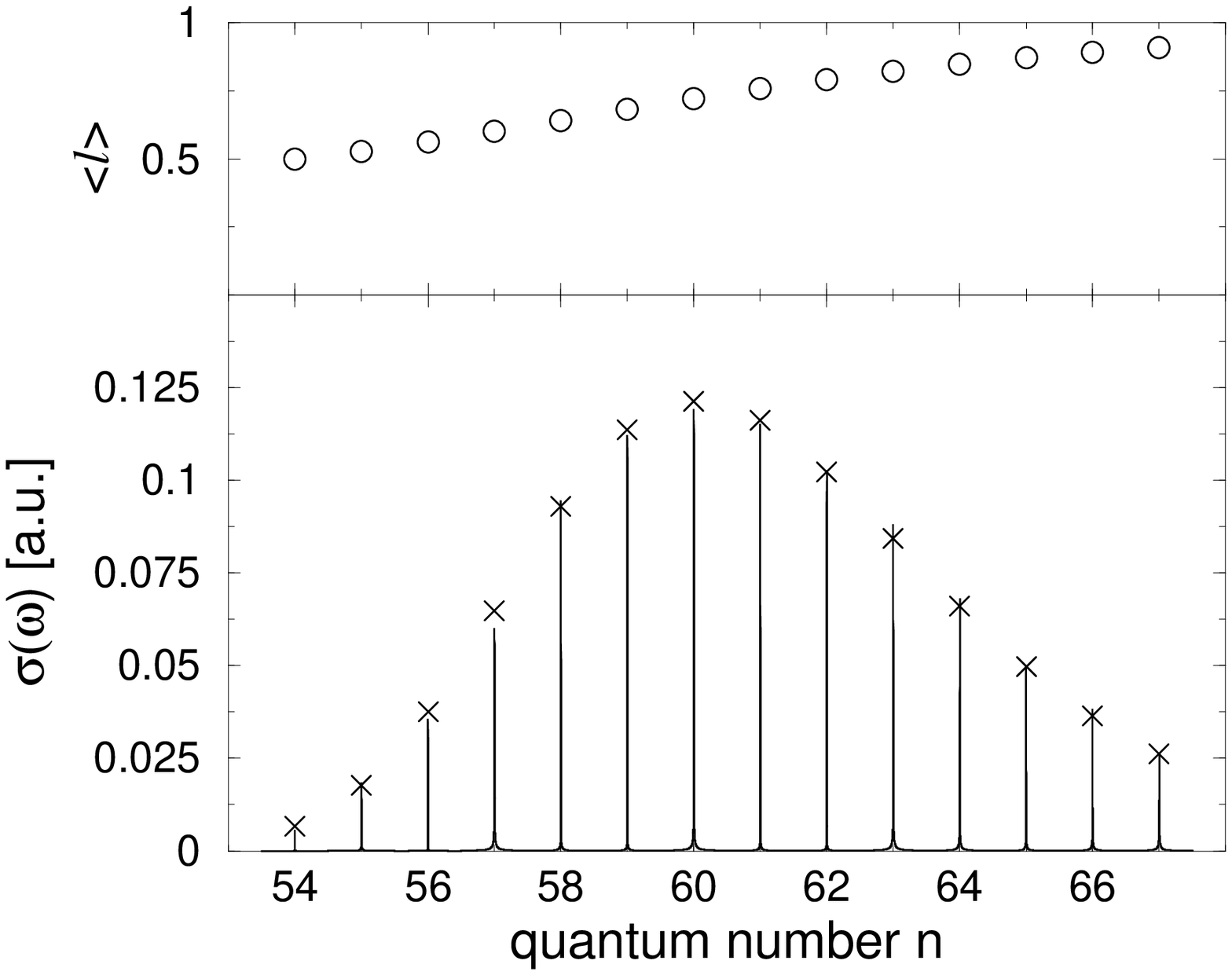}
\caption[]{Lower panel: Semiclassical spectrum (full curve) compared to
quantum spectrum (crosses) with parameters 
${\bf r}_0 = (0, 0, 6000 \, a.u.)$, ${\bf p}_0 = (0, 0.0075, 0)$ and $\gamma^2 = 2/600^2$; upper panel: average
normalized angular momentum according to \eq{lnorm} (circles).} \label{good}
\end{figure}

To support this finding we have prepared a different wavepacket with
an additional kick (initial momentum) perpendicularly to the axis 
connecting the center of the wavepacket and the Coulomb center. 
This creates a large fraction of high angular momentum states as can
be seen in \fig{good}. The agreement with the quantum power spectrum is in 
this case, covering the same energy window as \cite{Tom93,Ezr94}, and 
\fig{swave}, much better. Naturally, the one dimensional radial 
calculations of \cite{Tom93,Ezr94} have only $l=0$ states and in \fig{swave} 
the average angular momentum $\ell$ is also low by construction 
through the initial state. 

Hence, we can conclude that the power spectrum of hydrogen, including
the weights, can be reproduced semiclassically. While the semiclassical
energies $E_n$ are generally in good agreement with the quantum eigenvalues
the semiclassical weights $b_n$ are only accurate in the limit
of large quantum numbers, i.e. if the initial wave packet contains a
large fraction of high angular momentum states in each degenerate 
manifold $n$. This reflects the larger sensitivity of the weights described
by off--diagonal matrix elements, compared to the (diagonal) energies.
Seen in a wider context, our result implies the consequence that a one 
dimensional {\em radial} quantum problem is not really one dimensional. 
Rather, it is the limit of angular momentum $l = 0$ in three, or at least 
two dimensions. Hence, even for large quantum numbers $n$ in the radial problem 
the semiclassical limit is not reached since the angular momentum 
quantum number is zero.  The incomplete semiclassical limit causes 
in the case of the hydrogen problem the remaining discrepancies in the
purely radial semiclassical spectrum compared with the quantum spectrum. 
One may also view the failure of the one dimensional radial 
WKB treatment for $l = 0$ even for large quantum numbers $n$ as a 
consequence of this incomplete semiclassical limit.

In summary,  constructing the time correlation function semiclassically
in three cartesian dimensions with the help of the Herman-Kluk 
propagator we have demonstrated that the singular Coulomb potential  can be 
treated as any other non-singular interaction without any special 
precautions. Moreover, by virtue of our three-dimensional treatment, 
we could clarify the origin of the  discrepancies between the
quantum and the semiclassical calculation restricted to the radial 
dynamics only. We hope that this result stimulates future  
applications of semiclassical propagator techniques to Coulomb problems. 

We would like to thank Frank Gro\ss mann for helpful discussions on 
semiclassical initial value methods. JMR acknowledges the hospitality
of the Insitute for Advanced Study, Berlin, where part of this work 
has been completed.

This work has been supported by the DFG within the Gerhard Hess-Programm.


\begin{references}   
\bibitem{Kay94b} K.\ G.\ Kay, J.\ Chem.\ Phys.\ {\bf 101}, 2250 (1994).
\bibitem{Gro96} F.\ Gro\ss mann, Chem.\  Phys.\ Lett.\ {\bf 262}, 470 (1996).
\bibitem{Mil96} B.\ W.\ Spath and W.\ H.\ Miller, Chem.\  Phys.\ Lett.\ {\bf 262}, 486 (1996).
\bibitem{Tan96} S.\ Garashchuk and D.\ Tannor, Chem.\ Phys.\ Lett.\ {\bf 262}, 477 (1996).
\bibitem{HK84} M.\ F.\ Herman and E.\ Kluk, Chem.\ Phys.\ {\bf 91}, 27 (1984).
\bibitem{Kay94a} K.\ G.\ Kay, J.\ Chem.\ Phys.\ {\bf 100}, 4377 (1994).
\bibitem{Gro98} F.\ Gro\ss mann, Phys.\ Lett.\ {\bf A243}, 243 (1998).
\bibitem{Tom93} I.\ M.\ Su\'arez Barnes, M.\ Nauenberg, M.\ Nockleby and S.\ Tomsovic, 
Phys.\ Rev.\ Lett.\ {\bf 71}, 1961 (1993);  J.\ Phys.\ A {\bf 27}, 3299 (1994).
\bibitem{Ezr94} R.\ S.\ Manning and G.\ S.\ Ezra, Phys.\ Rev.\ A {\bf 50}, 954 (1994).
\bibitem{WN95} M.\ R.\ Wall and D.\ Neuhauser, J.\ Chem.\ Phys.\ {\bf 102}, 8011 (1995).
\bibitem{MT97} V.\ A.\ Mandelshtam and H.\ S.\ Taylor, Phys.\ Rev.\ Lett.\ {\bf 78}, 3274 (1997); 
J.\ Chem.\ Phys.\ {\bf 107}, 6756 (1997).
\bibitem{Klei95}  H.\ Kleinert,  {\it Path integrals
in quantum mechanics, statistics, and polymer physics}, 2nd ed., 
 (World Scientific, Singapore, 1995).
\end{references}
\end{document}